\documentclass[prl,aps,twocolumn,groupedaddress,showpacs,floatfix,preprintnumbers,superscriptaddress]{revtex4-1}
\usepackage{amsmath,amssymb,multirow,bm}

\newcommand{\beq}{\begin{equation}}
\newcommand{\eeq}{\end{equation}}
\newcommand{\bea}{\begin{eqnarray}}
\newcommand{\eea}{\end{eqnarray}}
\usepackage{graphicx}
\preprint{}

\begin{document}


\title{Big-bang nucleosynthesis constraints on dark-matter sectors revisited}
\author{C.A. Bertulani}
\email{carlos.bertulani@tamuc.edu}
\affiliation{Department of Physics and Astronomy, Texas A \& M University-Commerce, Commerce, Texas 75429, USA}
\author{V. Challa}
\email{vchalla1@leomail.tamuc.edu}
\affiliation{Department of Physics and Astronomy, Texas A \& M University-Commerce, Commerce, Texas 75429, USA}
\author{J.J. He}
\email{hejianjun@nao.cas.cn}
\affiliation{Key Laboratory of Optical Astronomy, National Astronomical Observatories, Chinese Academy of Sciences, Beijing 100012, China}
\affiliation{Key Laboratory of High Precision Nuclear Spectroscopy, Institute of Modern Physics, Chinese Academy of Sciences, Lanzhou 730000, China}
\author{S.Q. Hou}
\email{sqhou@impcas.ac.cn}
\affiliation{Department of Physics and Astronomy, Texas A \& M University-Commerce, Commerce, Texas 75429, USA}
\affiliation{Key Laboratory of High Precision Nuclear Spectroscopy, Institute of Modern Physics, Chinese Academy of Sciences, Lanzhou 730000, China}
\author{Ravinder Kumar}
\email{ravinderkumar12@yahoo.co.in}
\affiliation{Department of Physics and Astronomy, Texas A \& M University-Commerce, Commerce, Texas 75429, USA}
\affiliation{Department of Physics, Deenbandhu Chhotu Ram University of Science and Technology, Murthal 131039, Haryana, India}
\date{\today}

\begin{abstract}

We extend previous studies of big bang nucleosynthesis, with the assumption that ordinary matter and dark matter sectors are entangled through the number of degrees of freedom entering the Friedmann equations. This conjecture allows us to find a relation between the temperatures of the weakly interacting matter and dark-matter sectors. The constraints imposed by observations are studied by comparison with calculations of big bang nucleosynthesis for the abundance of light elements. It is found that these constraints are compatible with cold dark matter and a wide range number of dark sectors. 

\end{abstract}


\maketitle

\section{Introduction}

There is an overwhelming evidence that the Universe is dominated by dark energy and dark matter, with the visible matter amounting to only 5\% of its total energy/matter content. Dark energy takes about 70\% of it, while dark matter is responsible for the other 25\%. The evidences for the existence of Dark Matter (DM) are  based on galaxy rotation curves \cite{CorbelliEdvigeSalucciPaolo}, gravitational lensing \cite{doi:10.1146/annurev.astro.41.111302.102207,refId0}, and other recently observed phenomena \cite{Nature:Genzel}. The only property we know so far about DM is that it interacts gravitationally, but there is a growing consensus in the scientific community that it might consist  of a new particle beyond the standard model of particle physics.  The most popular candidate for this particle is the WIMP (Weakly Interacting Massive Particle), whose properties are basically unknown and left to wide theoretical speculation. Direct detection of such particles based on WIMP scattering induced nuclear recoils  have been pursued experimentally, but have yielded negative results so far (see, e.g., Ref. \cite{doi:10.1142/S0217751X15300380}).

Indirect searches for DM include looking for gamma-rays resulting from possible DM-antiDM annihilation or their decay (if they are unstable) into standard model particles. Such experiments are difficult because of the background from other typical astrophysical processes leading the same final channels \cite{doi:10.1142/S0217732305017391}. A far more indirect method is to use well established cosmological models whose predictions can be appreciably altered with the existence of DM.  One of such models is the Big Bang Nucleosynthesis (BBN) model, which has been proven to  successfully  describe the observed abundances of light elements, up to about nuclear masses $A\sim 7$, and trace them back to the primordial epochs of the Universe  \cite{RN312}.   The standard BBN predictions  are in a good agreement with the observed abundances of light elements. Deviations from these predictions can been used  as a test of new physics. The BBN predicted $^4$He abundance has become a benchmark for tests of big bang scenarios, although the BBN predicted abundances of deuterium and $^3$He are also in good agreement with observations. One exception seems to be the BBN predicted abundance of the $^7$Li isotope which is about a factor  3 larger than the observed values, under reasonable assumptions concerning astration and other possible lithium destruction scenarios.  The mismatch between BBN predictions and observations for $^7$Li has become known as the ``lithium problem".

Several solutions of the lithium problem have been proposed with physics ranging from the destruction of $^7$Be production during the BBN, to the influence of axion particles, and other ideas \cite{RN312}. Destruction of $^7$Be would possibly solve the lithium puzzle as the electron capture on $^7$Be is responsible for most of the $^7$Li produced in the BBN \cite{1475-7516-2012-06-030}. Recent analyses of the experimental data seem to rule out this possibility \cite{RevModPhys.88.015004,RN143}. Many other speculations such as the influence of non-standard  model particles in the BBN \cite{PhysRevLett.114.091101}, a hybrid model of axion dark matter \cite{Kusakabe2013704}, or variations of fundamental constants \cite{Berengut2010114} have been proposed.   Other effects which have been explored, such as the impact of electron screening on nuclear reactions during the BBN have been shown to be ineffective \cite{RN136}. More recently, the use of non-extensive statistics and its modification \cite{RN148} of the usually adopted Maxwell-Boltzmann distribution for the relative velocities of the particles in the plasma has proven to be a possible path to solve the lithium puzzle because it reduces the $^7$Be production without affecting the deuterium and helium abundances \cite{RN322} (see also \cite{AAS20171}). 

The influence of dark matter during the BBN has been studied in several publications (see, e.g., Refs. \cite{Dar:1995gh,1367-2630-11-10-105028,PhysRevD.65.123503,1475-7516-2013-02-010,Foot2012238,PhysRevLett.105.211304,Blum201430,Kawasaki2015246}).  In particular, the gauge model developed in Ref. \cite{BJPOliveira2016} proposes that the degrees of freedom associated with dark matter can increase the expansion rate of the early Universe. Analogous assumptions have  been used in Refs. \cite{BEREZHIANI199626,PhysRevLett.105.211304,PhysRevD.82.056001,PhysRevD.82.123512,Buckley2011}, the major difference being the number of dark sectors, which in Ref. \cite{BJPOliveira2016} was taken as five sectors instead of the usual one. The dark sectors could consist of parallel universes of dark matter interacting only via gravitation, but possibly being formed by identical copies of particles from our ordinary universe (i.e., dark photons, dark electrons, etc.).  A consequence of this conjecture is that it implies  an effectively larger number of degrees of freedom, as shown below.  

The influence of the degrees of freedom of the dark sectors straightforwardly arises from Friedmann equation for a flat Universe ($k=0$) and the absence of a cosmological constant, i.e. from the relation $H=\dot{R}/R=\sqrt{8\pi G\rho/3}$. The ratio of expansion of the Universe is proportional to the square root of the energy/matter density which obviously includes dark matter particles. For the typical temperatures during the BBN the light particles are relativistic and their contribution to the energy density is proportional to the number of their degrees of freedom entering the Stefan-Boltzmann $T^4$ law dependence on the temperature, and similarly with the entropy density $T^3$ dependence on the temperature. It is not necessary that the temperatures of ordinary matter and those in the dark sectors are equal and the question is how different they could be.  The standard BBN model assumes that the effective number of degrees of freedom during the early universe is $g_* = 10.75$, due to  photons,  $e^+e^-$ pairs, and the three neutrino species: $\nu_e$, $\nu_\mu$,  and $\nu_\tau$. 

The model developed in Ref.  \cite{BJPOliveira2016}  assumes that the mirror (dark) sectors are composed of particles which are exact copies of the standard model, but interacting weakly within and across all dark sectors and between the dark sectors and the ordinary one.  The model was shown to be compatible with the lepton anomalous magnetic moment (see also Refs. \cite{McKeen20111501,Chen201168}), with the muon beta-decay and with the B and D leptonic decays. It predicts an upper-bound for the dark proton-ordinary proton scattering cross section of the order of $2.5 \times 10^{-48}$ cm$^2$. As a consequence of the model, the number of degrees of freedom during the BBN increases. The presence of additional degrees of freedom of any relativistic species at $T_W \sim 1$ MeV (the ``freeze-out" temperature of weak interactions) increases $g_*$, and therefore the expansion rate, the value of $T_W$, the neutron to proton ratio, and finally the primordial elemental abundances.

Refs.  \cite{BEREZHIANI199626,BJPOliveira2016} have proposed that after inflation the temperature for the thermal baths associated with each particle species might not be the same. Such temperatures depend on the various possible reactions leading to equilibrium and on the thermal history of the Universe. The assumption used in Ref. \cite{BJPOliveira2016} relies on the simplest possible picture where all the five dark sectors have the same temperature, different from the ordinary matter thermal bath. This temperature difference might be due to an asymmetric reheating taking place after inflation, as envisaged in Refs. \cite{KolbSeckelNature1985,Berezhiani2001362,Ciarcelluti2008}. But the mirror model suggested in those works aimed at explaining the neutrino anomalies, and include mirror baryons about 20 times heavier than the familiar baryons. Such baryons play the role of the cold dark matter and they provide a reasonable explanation of why $\Omega_{DM}$ is of the same order as $\Omega_{Baryons}$.  In this work we will focus on the influence of the number of dark sectors in BBN and look for the constraints on such a number to test the hypothesis taken in Ref. \cite{BJPOliveira2016} on the existence of five parallel universes of dark matter. We show that whereas the temperature of the dark sectors can be constrained relatively well  with a  comparison with observations of the abundance of light elements, the number of dark sectors can vary wildly.

\section{Degrees of freedom in dark and ordinary matter sectors}

The most constraining parameters of BBN are the $^4$He primordial abundance and the baryon-to-photon ratio $\eta = n_b/n_\gamma$, where $n_b$ is the baryon density and $n_\gamma$ the photon density in the Universe \cite{Lisi.PhysRevD.59.123520}. They would be the best way to identify the number of possible new particles contributing to the radiation density during the BBN epoch. At very high temperatures, the energy content of the Universe was dominated by  radiation, the  particles (except for baryonic matter, i.e., protons and neutrons) were all relativistic and their masses can be neglected accordingly. In such a scenario the energy and entropy densities in this epoch are given by \cite{OlivePDG2010} ($\hbar=c=k_B = 1$)
\begin{equation}
 \rho(T)=\frac{\pi^{2}}{30} \, g_*(T) \,T^4 \quad\mbox{and}\quad
 s(T)=\frac{2\pi^{2}}{45} \, g_s(T)\, T^3 ,
 \label{energy_entropy}
\end{equation}
where
\begin{equation}
g_*(T)=\sum_B g_B \left(\frac{T_{B}}{T}\right)^4 + \frac{7}{8} \sum_F g_F \left(\frac{T_{F}}{T}\right)^4
\end{equation}
and
\begin{equation}
g_s(T)=\sum_B g_B \left(\frac{T_{B}}{T}\right)^3 + \frac{7}{8} \sum_F g_F \left(\frac{T_{F}}{T}\right)^3
\end{equation}
are the effective numbers of degrees of freedom during big bang nucleosynthesis. In these equations, $g_{B(F)}$ is the number of degrees of freedom of bosons (fermions) $B(F)$. The temperature of the thermal bath of each species are denoted by  $T_{B(F)}$, respectively, while  $T$ denotes the temperature of the photon thermal bath.

In Ref. \cite{BJPOliveira2016} a gauge model for the dark mirror sectors was introduced and several predictions for the scattering of dark sector particles with ordinary matter, mediated by the exchange of a Weakly Interacting Matter Gray (WIMG) boson with mass $M\sim 10 $ TeV,  was obtained. In particular, the elastic scattering of dark protons on ordinary protons was shown to be  $\sigma(p_dp) \simeq (m_N^2/64\pi)(g_M/M)^4$, for a nucleon mass $m_N = 1$ GeV and $g_M/M \sim 10^{-3}$ GeV$^{-1}$, yields $\sigma(p_dp)< 10^{-48}$ cm$^2$. Therefore, thermalization involving dark mater and ordinary matter must have decoupled much before BBN ensued, even if the coupling constant $g_M$ would have increased dramatically with increasing densities and temperatures. If the dark sectors are composed of particles which are exact copies of the ordinary sector, then the dark photons, dark $e^{\pm}$ and dark neutrinos were also relativistic before decoupling. After the matter-DM decoupling, the temperatures decreased simultaneously as the Universe expanded, $T \sim 1/R$, and the expansion rate dependence with time was $R \sim t^{1/2}$ according to Friedmann equations. Hence, it is fair to assume that during the BBN the matter and dark matter sectors are decoupled, with different temperatures: $T$ for matter and $T'$ for the dark matter bath. During this epoch and later on, the remaining interaction between all sectors is overwhelmingly due to the gravitational force manifest through the respective densities $\rho$ and $\rho'$ of matter and DM entering the Friedmann equations. 

For the dark matter bath, the energy $\rho'(T')$ and entropy $s'(T')$ densities are given as in Eq. \eqref{energy_entropy} but with the effective number of degrees of freedom changed to $g_*(T) \rightarrow g'_*(T')$ and $g_s(T) \rightarrow g'_s(T')$, and replacing $T$ by $T'$. If  entropy in each sector is separately conserved during expansion, then $x=(s'/s)^{1/3}$ is time independent. In this case, unless the temperature of the dark sector $T'$ is much smaller than that of ordinary matter, with the same relativistic particle content for each sector one would reach to the conclusion that the initial condition  $g_s(T_0) = g_s^\prime(T'_0)$ implies that  $x=T'/T$.   It is not ruled out that the dark matter particles might self interact  via some unknown force besides gravity. Even if it is  small compared to the known interactions, the assumption that their masses are the same as particles in the visible sector, with same kinds of dark matter particles, and same status as fermions or bosons,  does not imply that their mean velocity  and degree of freedom have to be shared among all of them equally as the Universe  expands. Here we do not speculate based on such unknown facts.

Cosmological models have been proposed with self-interacting cold dark matter particles  with a large scattering cross section but negligible annihilation or dissipation \cite{PhysRevLett.84.3760}.  Present observations of the dark matter halos seem to suggest that self-interacting dark matter particles have an upper bound for their cross section of the order of ${\sigma_x/ m_x} \sim 10^{-25}$ cm$^{-2}$ GeV$^{-1}$, where $\sigma_x$ is the cross section and $m_x$ the mass of the dark matter particle \cite{Wandelt2000,Escude2000,Dave2001,Yoshida2000}. Cross sections of weakly interacting particles with energy $E$ in a thermal bath with temperature $T$ are typically given by $\sigma \approx g^2E^2\approx g^2 T^2$, where $g$ is the interaction coupling constant and $T$ is the temperature. Assuming  that $g$ is of the same order of magnitude in the dark sectors, one has $\sigma'/\sigma = (T'/T)^2$, where $\sigma'$ ($\sigma$) is the cross section for the dark (ordinary) particles, e.g. pp or p$_d$p$_d$ scattering. Thus, unless $T^\prime \ll T$, the magnitude of the elastic cross sections in the dark sectors would be of similar magnitude as with the ordinary matter, but with a tendency of dark sectors being less collision intense. 

As we show next, detailed BBN calculations for light element abundances do not disallow these conclusions when BBN predictions are compared to observations.  

\section{Nucleosynthesis constraints on the temperatures and number of dark sectors} 

During the BBN and temperatures of the order of 1 MeV, photons,  $e^+e^-$ pairs, and three neutrino species $\nu_e$, $\nu_\mu$,  and $\nu_\tau$ are in a quasi-equilibrium. The number of degrees of freedom at $m_e < T <m_\mu$ is $g_{*}(T)|_{T=1MeV} = 43/4=10.75$ \cite{weinberg2008cosmology}. 
Including an yet to be determined number of dark sectors $N_{DM}$, the Friedman equation during the radiation dominated era is given by
\begin{equation}
H(t)=\sqrt{\left(8\pi/3 c^2\right) \, G_{N} \, \bar{\rho}}.
\end{equation}
where the comoving energy density is $\bar{\rho} = \rho\, + \, N_{DM} \, \rho'$, with $\rho$ being the energy density of ordinary matter and  $\rho^\prime$ the energy density in the dark matter sectors. Using Eq. \eqref{energy_entropy} for $\rho$, and $\rho'$ leads to
\begin{equation}
H(t)=\sqrt{{8\pi^3\over 90}\bar{g}_{*}(T)}   \frac{T^2}{M_{Pl}},
\end{equation}
where  $M_{Pl}=\sqrt{\hbar c^5/G}$ is the Planck mass and
\begin{equation}
\bar{g}_{*}(T) = g_{*} (T) \left( 1+ N_{DM} \, \xi \, x^4 \right). \label{newdof}
\end{equation}
The parameter $\xi$ is given by $\xi = \left(g'_{*}/g_{*}\right) \left(g_{s}/g'_{s}\right)^{4/3}$ and is equal to 1, unless $T'/T$ is very small, following the same arguments given before. 
Therefore, the number of the effective degrees of freedom  during the BBN in the presence of $N_{DM}$ dark sectors change from $g_{*}$ to $\bar{g}_{*} = g_{*} \left(1 + N_{DM} \, x^4 \right)$.  

\begin{figure}[ptb]
\begin{center}
\includegraphics[
width=3.4in]{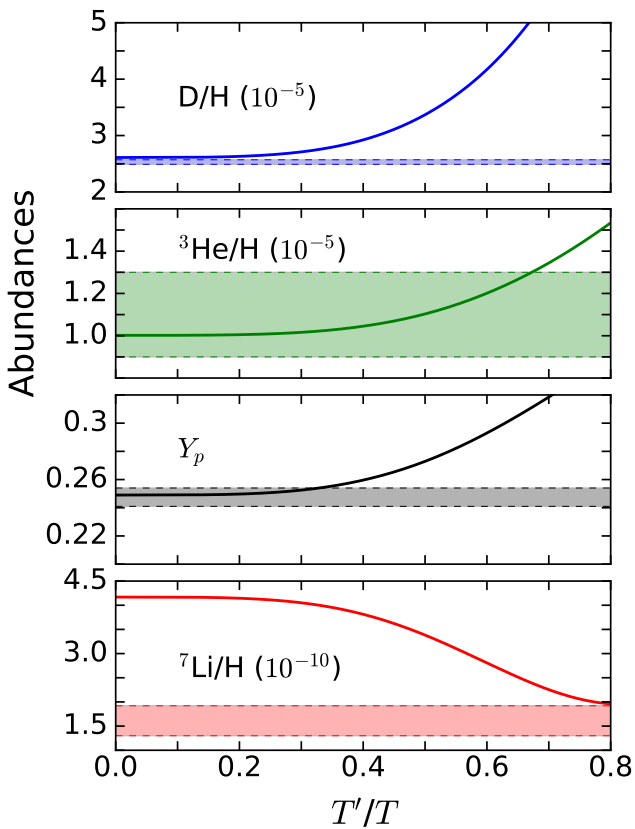}
\end{center}
\caption{BBN prediction for the relative abundances of D, $^3$He, $^4$He (mass fraction, $Y_p$) and $^7$Li as a function of the ratio of the temperature of dark sectors to that of the ordinary matter. Here we assume that the number of dark sectors is fixed to $N_{DM}=5$. The shaded bands represent the uncertainty in the observed values.}
\label{fig1}
\end{figure}

The bounds for $N_{DM}$ and $x$ can be assessed by the relative abundances of the light element isotopes (D, $^{3}$He, $^{4}$He, and $^{7}$Li).  In Figure \ref{fig1} we show the BBN prediction for the relative abundances of D, $^3$He, $^4$He and $^7$Li as a function of the ratio of the temperature of dark sectors $T'$ to that of the ordinary matter, $T$. Here we assume that the number of dark sectors is fixed to $N_{DM}=5$. Our calculations for the primordial abundances are performed with a standard BBN code, with a baryon to photon ratio $ \eta =6.2 \times 10^{-10}$, the number of neutrino families $N_\nu=3$ and the neutron lifetime $\tau_n=880.3$ s \cite{OlivePDG2010}. For simplicity, we do not include experimental errors associated with input quantities to calculate the elemental abundances. Therefore, the shaded bands represent the uncertainty in the observed values. 

The usually most stringent constraint, namely the abundance of helium, shows that the dark sectors should have a temperature smaller than the ordinary matter, with $T'\lesssim 0.3 T$. This is  also partially supported by the relative abundance of  deuterium, while the observations on the relative abundance of $^3$He allow for a larger temperature for the dark sectors \cite{doi:10.7566/JPSCP.14.010102}. Notice that the uncertainty in the deuterium abundance has been dramatically decreased in recent observations \cite{0004-637X-781-1-31,doi:10.7566/JPSCP.14.010102} and, if taken at face value, would imply a smaller value of $T'/T$.  
Interestingly, the $^7$Li relative abundance is better described by a larger temperature for the dark sectors. This result is rather tempting, due to a possible solution of the so-called lithium puzzle by the addition of new degrees of freedom of particles in the dark sectors. Evidently, the constraints set by the  D, $^3$He, and $^4$He do not allow for this possibility, at least under the scenario that all abundances are related by the same BBN reaction network, with or without dark sectors. We thus conclude that the dark sectors must be cold compared to the ordinary matter during the BBN.

We notice from Eq. \eqref{newdof} that the value of $\bar{g}_{*}(T) $ is much more sensitive to $T'$ than it is to $N_{DM}$. In fact, if we set $T'=0.3T$ as a typical temperature value for cold sectors, we find that here is a large range of  $N_{DM}=1-50$ for which the relative abundances of D, $^3$He and $^4$He  would fall within the constraints set by observation. It is worthwhile mentioning that recent observations have yielded a much smaller uncertainty on the primordial deuterium abundance. We also observe that only if we assume a value $N_{DM}\simeq 100$ would the observed abundance of $^7$Li be explained by BBN theory. However, as one infers from the extrapolation of the solid curves on the far right  of Figure \ref{fig1}, the BBN predictions of   D, $^3$He and $^4$He would fall way off the observations.

Another way to cast the dependence of BBN predictions on the extra degrees of freedoms introduced by the existence of dark sectors is to add the uncertainty in the number of massless neutrinos during nucleosynthesis, which is supposed to be in the interval $ 2.92 <  N_{\nu} < 3.38 \ (1\sigma)$ \cite{refId0}, in agreement with the predictions of the Standard Model (SM). Other probes, with less resolution than the Planck satellite, set this uncertainty within the interval $3.46 <  N_{\nu} < 5.2$ \cite{0067-0049-192-2-18}. Here we will use Planck results to study the  impact of the neutrino families on constraining the number and temperature of dark sectors. 

Deviations from the accepted value, $g_* = 10.75$, are often expressed in terms of the number of extra neutrino species, $\bar{g_*} = g_* -10.75 = 1.75 \Delta N_\nu$. We introduce a  parametrization in which the extra degrees of freedom due to the dark sectors and the number of neutrino families is expressed as
\begin{equation}
\bar{g}_{*}=g_{*} \left[1+   1.75 \Delta N_\nu + N_{DM} \, \xi \, x^4 (1+1.75 \Delta N^{dark}_\nu) \right],
\end{equation}
where $\Delta N_\nu$ is the variation in equivalent number of neutrinos. That is, we include degrees of freedom due to additional neutrinos species, and we further assume for simplicity that $N_\nu=N^{dark}_\nu$. Based on our previous conclusions, we assume $T'=0.3T$   to reconcile with the BBN data with $N_\nu =3$. We also set  $N_{DM} = 5$, as required to explain the observed ratio $\Omega_{DM} / \Omega_b$. In the standard BBN model neutrinos decouple from other particles at a temperature of  $\sim 2 - 3$ MeV before electron-positron pairs annihilate. Therefore, they do not share the entropy transfer from the $e^\pm$ pairs. That is why the neutrino temperature is smaller than the photon temperature. Our model does not imply that dark matter couples with neutrinos, a popular approach in recent works to account for light ($M \lesssim 20$ MeV) dark matter particles coupling to  neutrinos via dark-matter annihilation \cite{Heo2016}. In these models, the  neutrino-to-photon temperature ratio can change, as well as  the effective number of neutrinos degrees of freedom. 

\begin{figure}[ptb]
\begin{center}
\includegraphics[
width=3.4in]{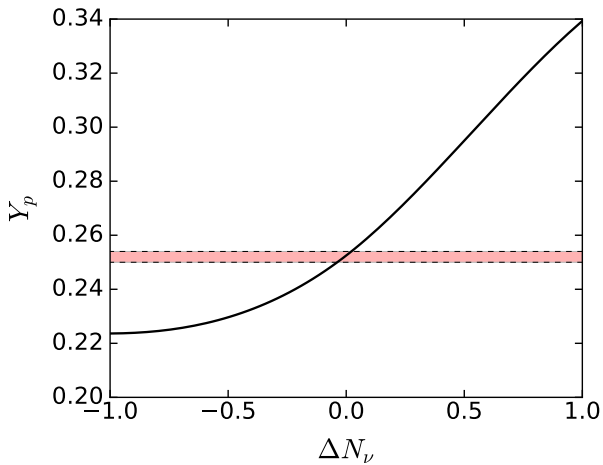}
\end{center}
\caption{Dependence of BBN prediction for the primordial $^4$He  as a function of the additional number of neutrino families, including the impact of the degrees of freedom of particles in dark sectors with a temperature $T'= 0.34T_{BBN}$ and the number of dark sectors $N_{DM}=5$.}
\label{fig2}
\end{figure}

In Figure \ref{fig2} we show the dependence of BBN prediction for the primordial $^4$He  as a function of the additional number of neutrino families, including the impact of the degrees of freedom of particles in dark sectors with a temperature $T'= 0.34T_{BBN}$ and the number of dark sectors $N_{DM}=5$. One sees that the estimate of number of dark sectors and their temperature does not affect the constraints set by observations (shaded area), unless the number of neutrino families is decreased ($\Delta N_\nu <0$) or increased   ($\Delta N_\nu >0$) substantially from the expected value of $N_\nu =3$.

The baryon asymmetry is parameterized by the baryon-to-photon ratio $\eta$.
The density number of photons $n_{\gamma}$ is proportional to T$^{3}$ and, therefore,
one can write the density number of dark-photons as $n'_{\gamma}=x^3 n_{\gamma}$. The
ratio of dark-baryons to ordinary-baryons is given by $\beta=\Omega'_{b}/\Omega_{b} = x^3 \eta'/\eta$
\cite{Berezhiani2001362}. The bounds from the BBN on $x = T' / T$ imply that the baryon asymmetry in the
dark sector is greater than in the ordinary one. Indeed, using as an upper bound the estimate $x\sim 0.8/N^{1/4}_{DM}$ 
and assuming that each sector contributes equally to the Universe's energy
density $\beta\sim 1$, we obtain $\eta' \sim 2.1 \, N^{3/4}_{DM}\eta$, in agreement with Ref. \cite{BJPOliveira2016}. For the special where
$N_{DM} = 5$ it follows that $\eta'\sim 7\eta$. Asymmetric dark matter models, see e.g.
\cite{PhysRevLett.105.211304,PhysRevD.82.056001,PhysRevD.82.123512,Buckley2011}, give similar results for the baryon asymmetry. Therefore, our assumptions are not farfetched as they might initially seem.

\begin{table}[htbp]
\vspace{0.0cm}
\centering
\caption{\label{tab:table1} Predictions of BBN calculations using $ \eta=6.2\times 10^{-10}$,  $N_\nu=3$ and $\tau_n=880.3$ s. The numerical values in the second (third) [fourth] columns assume $N_{DM}$ and $T'/T$ equal to 1 and 0.3 (5 and 0.3) [5 and 1], respectively. The last column are absevations reported by several groups.}
\begin{tabular}{|c|c|c|c|c|c|c}
\hline
\hline
$N_{DM}=$ &$1$&$5$ &$5$&Observations  \\ 
$T'/T=$ &$0.3$&$0.3$& $1$&\\
 \hline
$Y_p$&0.246&0.249&0.376&$0.2449 \pm 0.004$ \cite{1475-7516-2015-07-011} \\
D/H ($10^{-5}$) &2.57&2.62&10.1&$2.53 \pm 0.04$\cite{0004-637X-781-1-31}\\
${^3}$He/H ($10^{-5}$)&0.94&0.97&1.76&$(1.1\pm 0.2)$ \cite{Bania20002}\\
${^7}$Li/H ($10^{-10})$&4.41&4.38&1.72&$(1.58\pm 0.31)$ \cite{Sbordone2010}  \\ \hline
\hline
\end{tabular} 
\vspace{0.0cm}
\label{tab1}
\end{table}     

In Table \ref{tab1} we show our numerical predictions for the BBN abundances of light elements using the baryon-to-photon ratio $ \eta=6.2\times 10^{-10}$,  the number of neutrino families $N_\nu=3$, and the neutron lifetime $\tau_n=880.3$ s as input. The numerical values in the second (third) [fourth] columns assume $N_{DM}$ and $T'/T$ equal to 1 and 0.3 (5 and 0.3) [5 and 1], respectively. The last column are observations reported by several groups.   

\section{Conclusions}

In summary, we have studied effects of dark-matter degrees of freedom in the early Universe on the abundances of the light elements synthesized during the BBN epoch. Comparing our theoretical predictions  on the primordial abundances of the light elements with the observed values, we have obtained bounds for the possible number of dark matter sectors, $N_{DM}$. We show that the number of dark sectors is only loosely constrained by the comparison of BBN predictions with observations.  The possible value of $N_{DM} =5$, based on the observed ratio between dark matter and visible matter is a reasonable possibility. The temperature $T'$ in the dark sectors is shown to be colder than $T_{BBN}$ in the ordinary matter sector, albeit not much smaller: $T' \sim 0.3 T_{BBN}$.    

We have also verified that the number of neutrino families, $N_\nu = 3$  is compatible with the existence of  multiple dark sectors with a colder temperature.  The constraints set by the number of neutrino families do not change our predictions for $T'$ and $N_{DM}$.

\section*{Acknowledgement} 

The authors are indebted to W. de Paula, O. Oliveira and A. Coc for useful discussions. This work was supported in part by the U.S. DOE grants DE-FG02-08ER41533 and  the U.S. NSF Grant No. 1415656.   J.J. He and S.Q. Hou are supported by the the National Natural Science Foundation of China (Nos. 11490562, 11675229), and the Major State Basic Research Development Program of China (2016YFA0400503).



\end{document}